# Automated end-station at the soft X-ray beamline featuring robotic air-to-ultrahigh-vacuum sample transfer, scanning probe spectroscopy, and machine learning analysis

**Keita Hiromori[a‡*], Masashige Miyamoto[a‡], Masafumi Horio[a*], Takeru Nakashima[b], Yasunobu Ando[b], Masato Kotsugi[c], and Iwao Matsuda[a]**

[a] The Institute for Solid State Physics, The University of Tokyo, Kashiwa, Chiba 277-8581, Japan

[b] Institute of Integrated Research, Institute of Science Tokyo, Yokohama, Kanagawa 226-8501, Japan

[c] Department of Materials Science and Technology, Tokyo University of Science, Tokyo, 125-8585, Japan

‡ These authors contributed equally. * Correspondence e-mail: hiromori@issp.u-tokyo.ac.jp, mhorio@issp.u-tokyo.ac.jp

**Synopsis** We developed a scanning probe spectroscopy station at the soft X-ray beamline BL08U in NanoTerasu equipped with robotic air-to-ultrahigh-vacuum sample transfer, automated spectroscopic measurement, and various machine learning analysis capabilities. User experiments can be operated without any beamline staff, and the data can be curated remotely.

**Abstract** Driven by the growing integration of artificial intelligence and robotics in experimental science, we redefine the execution of user experiments at synchrotron soft X-ray beamlines by developing a highly automated end-station that can be operated without on-site staff. Installed at beamline BL08U of NanoTerasu, the system comprises four chamber units and two robots that automatically transfer samples from ambient air to an ultrahigh vacuum measurement stage for spectroscopic evaluations, before returning them to their original locations post-measurement. Soft X-ray absorption spectra are automatically acquired once programmed by scanning the micro-focused beam over the designated sample area, followed by automated evaluation using machine learning techniques such as principal-component analysis, multidimensional scaling, t-distributed stochastic neighbour embedding, and uniform manifold approximation and projection. The high-throughput data are systematically accumulated in the database for remote user examination. This unstaffed end-station provides efficient beamtime for users and allows beamline scientists/technicians to pursue more human and creative endeavors.

## 1. Introduction

Properties and functions of materials originate from their chemical composition, local structure, and electronic states. Reliable materials characterization is therefore essential not only for understanding existing materials but also for guiding the discovery and optimization of new ones. Soft X-ray absorption spectroscopy (XAS) is a powerful tool for this purpose because it probes element-specific excitations from core levels to unoccupied states, providing information on chemical states, orbital characters, and local electronic structures (Zimmermann *et al.*, 2020; Chantler *et al.*, 2024). Consequently, XAS has been widely applied to the characterization of catalysts, energy-storage materials, magnetic materials, and compounds containing light elements (Hiromori *et al.*, 2026; Wu *et al.*, 2021; Sumi *et al.*, 2022; Yamaguchi *et al.*, 2024, 2026).

Many functional materials contain particles, domains, interfaces, and chemically heterogeneous regions with characteristic dimensions in the micrometer and sub-micrometer ranges (Kondo, 2017; Sugawara *et al.*, 2023; Guan *et al.*, 2024; Zhang *et al.*, 2024; Hiromori *et al.*, 2024; Kawamura *et al.*, 2025; Miyamoto *et al.*, 2025; Horio *et al.*, 2025; Zhang *et al.*, 2025; Ozawa *et al.*, 2025). Their macroscopic properties are governed by these local structures, although the associated variations in electronic structure are obscured in spatially averaged spectra. Scanning soft XAS using a focused beam addresses this limitation by rastering a focused beam across a sample surface and recording an absorption-related signal as a function of position and photon energy.

The rapid development of artificial intelligence (AI) is transforming data analysis and knowledge discovery in materials research. Machine learning methods can extract latent spectral components, visualize relationships among spectra, and identify chemically distinct or anomalous regions that are difficult to recognize by manual inspection (Hasegawa *et al.*, 2025). These approaches are particularly effective for the large, multidimensional datasets generated by scanning spectroscopic methods, in which spectra are acquired across many spatial positions and experimental conditions. However, reliable and interpretable AI-assisted analysis requires sufficiently large datasets acquired under consistent and well-controlled conditions. The growing use of AI in materials research therefore creates a corresponding demand for experimental systems capable of producing high-quality spectroscopic data with substantially greater throughput.

Meeting this demand requires advances not only in data analysis but also in the hardware and control systems used for data generation. Robotics and automated control can increase experimental throughput by reducing repetitive operations and enabling extended measurements with minimal intervention from beamline staff. Automated operation has been demonstrated for hard XAS measurements (Mangold, 2018), tomographic microscopy (Mader *et al.*, 2011), total-scattering measurements (Yamada *et al.*, 2022), small-angle X-ray scattering (Lazo *et al.*, 2022), and powder diffraction (Stone *et al.*, 2023). Most existing implementations, however, employ hard X-rays, whose high transmission through air and window materials facilitates sample handling under atmospheric conditions. Conversely, soft X-ray experiments require vacuum beam paths and measurement chambers, thus requiring an automation architecture specifically designed for vacuum-compatible sample transfer and measurement.

To address this challenge, we developed the PIONEER (Process Informatics OperaNdo Experiment for Electronic Research) system at the NanoTerasu beamline BL08U. We combined our custom-built automated sample-handling and vacuum-control system with a previously developed scanning soft X-ray spectrometer (Ando *et al.*, 2023). Atmospheric and vacuum-compatible robots transfer sample holders from a five-position atmospheric sample bank, through a transfer chamber that alternates automatically between atmospheric and ultrahigh vacuum (UHV) conditions, to the measurement stage. Venting, evacuation, gate-valve operation, robotic transfer, scanning, data acquisition, data storage, and sample return are fully coordinated within a single workflow.

After the operator identifies a target sample, registers the spatial and photon-energy conditions, and initiates the measurement recipe, the PIONEER system executes the routine autonomously without requiring the continuous presence of on-site beamline staff. XAS data are transferred automatically to cloud storage, while the associated measurement conditions and beamline parameters are registered in a linked metadata database for remote examination. In this study, Si knife-edge measurements were used to evaluate the focused-beam size and transfer reproducibility, and Ni-mesh measurements demonstrated the integrated transfer, scanning, and XAS functionalities. A dataset of 121 Ni *L*-edge spectra acquired on an 11 × 11 spatial grids was subsequently analyzed using principal-component analysis (PCA), multidimensional scaling (MDS), t-distributed stochastic neighbor embedding (t-SNE), and uniform manifold approximation and projection (UMAP). Ultimately, the PIONEER system combines unstaffed soft X-ray measurements, large-scale dataset generation, automatic data registration, and online machine learning analysis.

## 2. Overall system and measurement workflow

The PIONEER system is installed at the soft X-ray beamline BL08U of NanoTerasu. BL08U is a high-brilliance soft X-ray beamline equipped with an APPLE-II-type undulator that generates linearly or circularly polarized soft X-rays. The beamline employs a collimated plane-grating monochromator comprising four mirrors (M0–M3) and a plane grating, covers a photon-energy range of 180–2500 eV, and provides an energy resolving power, $E/\Delta E$, exceeding 20,000. The detailed design of the BL08U optical system has been reported previously (Murano *et al.*, 2025). The system integrates an automated sample-handling and vacuum-control system with a scanning soft XAS chamber equipped with a Wolter mirror. The integrated system comprises an atmospheric sample bank housed in an acrylic box, a transfer chamber, a UFO chamber, a main chamber, atmospheric- and vacuum-compatible robots [Fig. 1(*a*)], and a supervisory control interface [Figs. 1(*b*) and (*c*)].

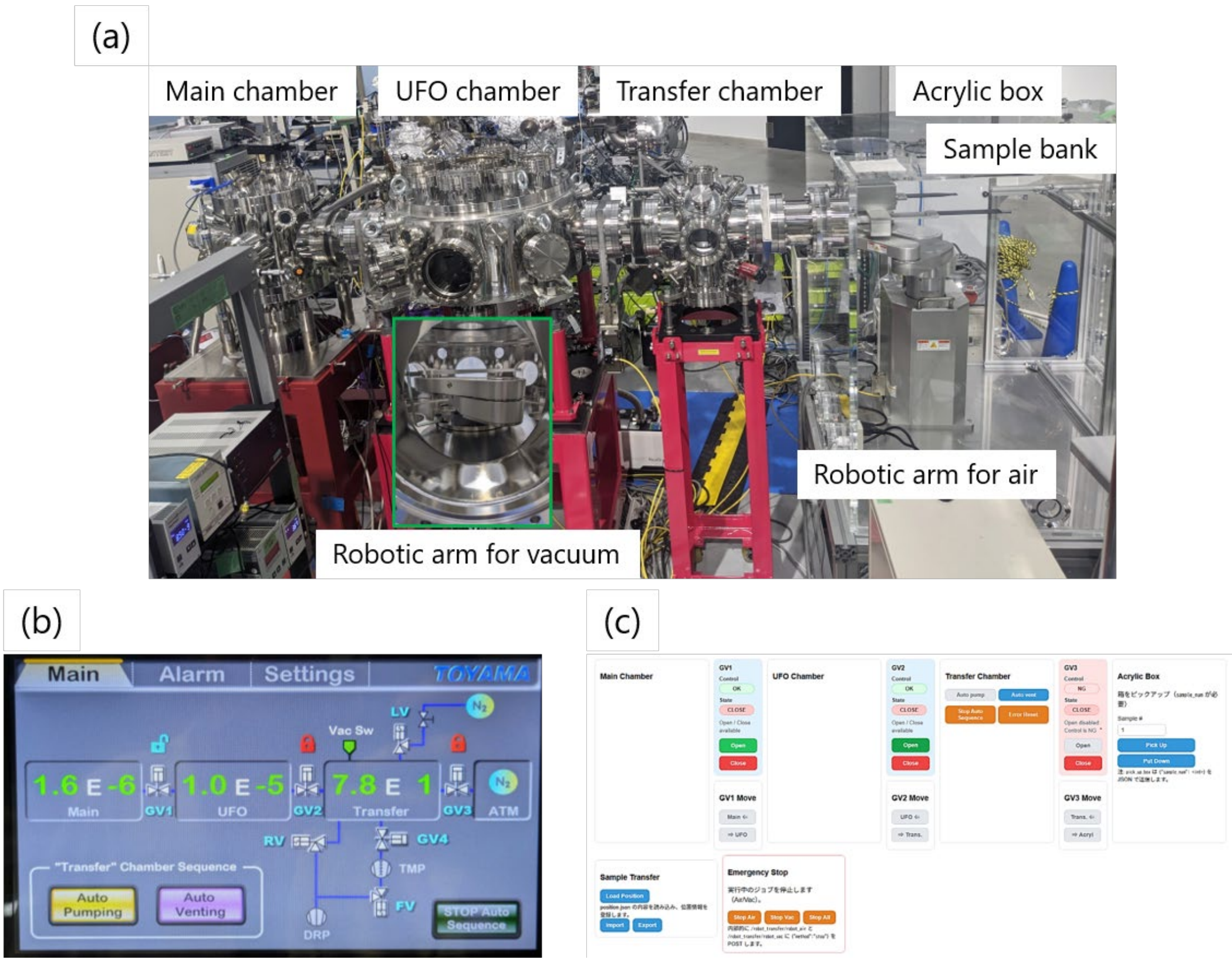

**Figure 1** Overview of the integrated system: (*a*) experimental apparatus with the atmospheric and vacuum robotic arms; (*b*) representative chamber-status readouts; and (*c*) supervisory interface for robots, pumps, gate valves, and measurement execution.

Fig. 2 schematically summarizes the unstaffed measurement and data-analysis workflow. The operator first loads the sample holders into the sample bank in the acrylic box. With the transfer chamber at atmospheric pressure, the atmospheric robot places the selected holder inside the transfer chamber. The atmospheric-side gate valve then closes, and the chamber is automatically evacuated. When the pressure reaches high vacuum (~$10^{-4}$ Pa) and the corresponding control flag is confirmed, the vacuum-side gate valve opens. The vacuum-compatible robot in the UFO chamber retrieves the holder from the transfer chamber and places it on the piezoelectric scanning stage in the main chamber. The tapered configuration guides the holder into its reproducible seating position without manual alignment.

After sample installation, the system executes the raster scan on the registered area automatically, recording the photon energy, stage coordinates, X-ray absorption signal, incident-photon-intensity monitor signal, and associated metadata (e.g., beamline parameters). The acquired XAS signal is normalized using the drain current measured from an upstream Au mesh. Measurement files are transferred automatically to cloud storage, while the corresponding acquisition conditions and beamline parameters are registered in a linked metadata database. The resulting spectroscopic dataset is subsequently analyzed via machine-learning pipelines, and the analysis data are stored in the data system, as indicated by the arrow in Fig. 2. Selected beamline specifications and principal operating values of the integrated system are summarized in Table S1 of the supporting information.

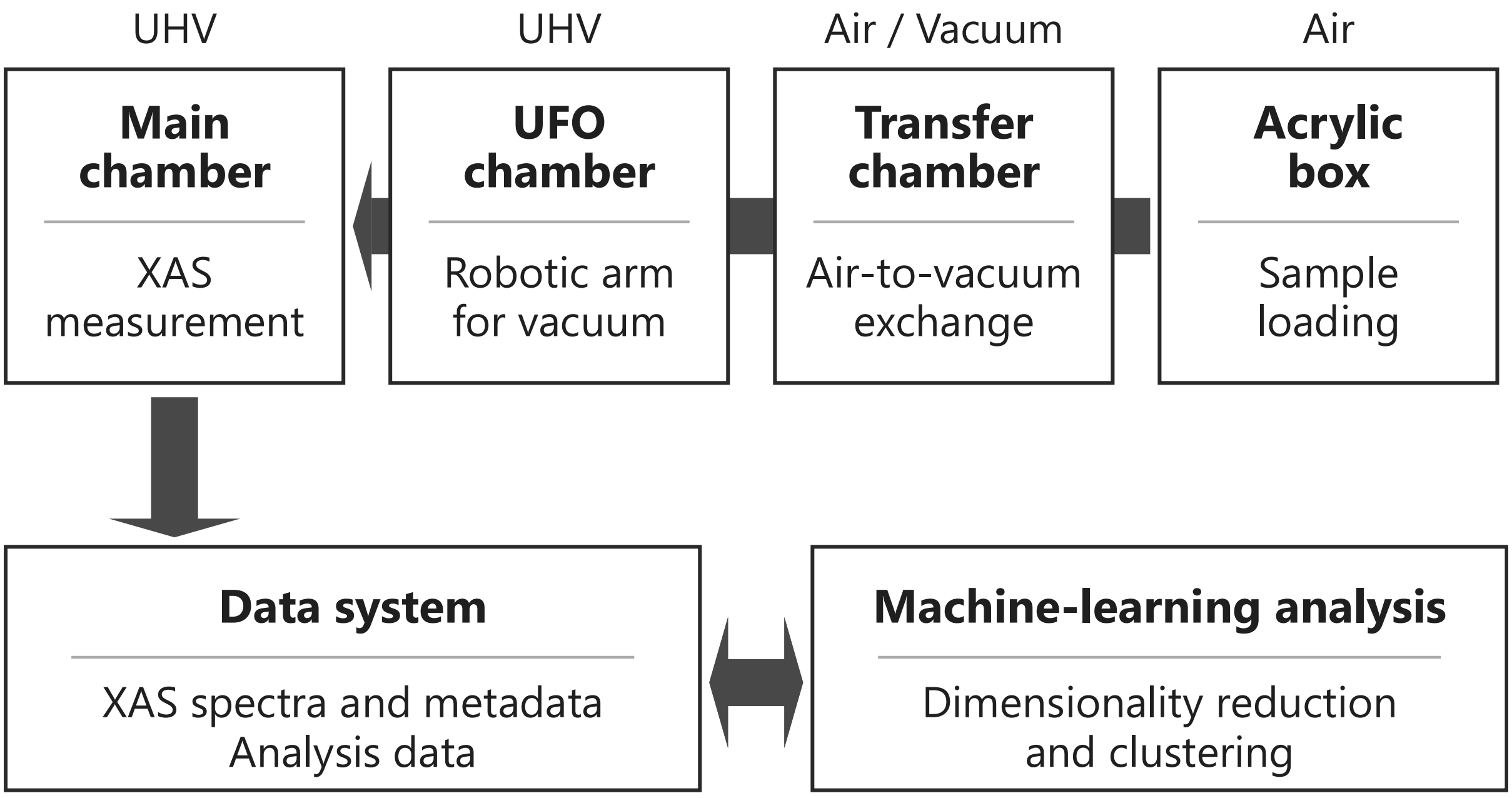

**Figure 2** Schematic overview of sample handling, XAS measurement and data analysis in the PIONEER system. A sample holder is loaded in the acrylic box, transferred through the transfer chamber during the air-to-vacuum transition, and carried by the vacuum-compatible robot in the UFO chamber to the main chamber for XAS measurement. The acquired spectra and measurement metadata are stored in the data system. The bidirectional arrow between the data system and machine-learning analysis indicates that stored data are used for dimensionality reduction and clustering, and that the resulting analysis data are registered in the data system.

### 2.1. Robotic air-to-UHV sample-transfer system

The sample bank is housed in the acrylic box and stores up to five sample holders. Additional sample-holder banks can be installed within the acrylic box, increasing the number of samples that can be queued for measurement. Sample transfer on the atmospheric side is performed by a JEL STCR4160S-300-PM robot controlled using a C4400 controller. The selected holder is moved from its registered position in the sample bank to the transfer chamber. After the transfer chamber is isolated and automatically evacuated to the transfer pressure, a vacuum-compatible JEL SVCR3190-040-PM robot, controlled using a C4300 controller, retrieves the holder and transfers it to the measurement stage in the main chamber. Following completion of the XAS measurement, the two robots execute the transfer sequence in reverse, returning the holder to its registered position in the atmospheric sample bank. The automated robotic sample-transfer sequence is demonstrated in Appendix A, Movie 1.

The transfer chamber serves as the physical interface between the atmospheric and vacuum environments. To allow access for the atmospheric robot, the chamber is automatically vented with $N_2$ gas. After the sample holder is placed inside, the chamber is isolated and automatically evacuated to a transfer pressure of approximately $10^{-4}$ Pa, enabling the vacuum-compatible robot to retrieve the holder and transfer it to the main chamber. A standard evacuation cycle requires approximately 3 min. The ultimate pressure of the transfer chamber is approximately $10^{-6}$ Pa, and the main chamber is maintained at approximately $10^{-6}$ Pa. Sample introduction and retrieval each require approximately 10 min, resulting in a total sample-exchange time of approximately 20 min.

The sample holder assembly consists of two primary components: an Al sample plate onto which the samples are mounted, and a SUS304 holder base incorporating a tapered positioning structure [Figs. 3(*a*) and (*b*)]. Depending on the mounting pattern, multiple samples can be accommodated on a single plate. Fig. 3(*c*) shows a representative mounting configuration, including a Si knife edge and several samples, along with the *X-Z* scan directions. Samples are secured to the Al plate using conductive carbon tape, and the plate is fixed to the SUS304 holder base using M3 bolts. The mounting geometry is designed to position the exposed sample surfaces at the nominal focal plane of the Wolter mirror.

The tapered guide surfaces of the SUS304 base constrain the lateral position and orientation of the holder assembly as it seats into the corresponding receiver [Fig. 3(*d*)]. Tapered receivers are installed within the sample bank, transfer chamber, and main chamber. The same mechanical interface is therefore used at every storage and transfer position, reducing positioning errors during successive robotic handoffs and improving the reproducibility of holder placement. The receiver in the main chamber is mechanically registered to the focus position of the Wolter mirror. Consequently, seating the holder in this receiver passively returns the sample surface to the nominal X-ray focal plane, providing a self-aligning function that eliminates the need for manual mechanical adjustment. The taper angle is 38°, defined as the acute angle between the inclined guide surface and the horizontal seating surface.

At the measurement position, the holder base is seated in a receiver mounted on two orthogonally arranged CLS-5252 piezo-driven linear stages with exchangeable drive units (SmarAct GmbH), as shown in Fig. 3(*e*). The stages are controlled using an MCS2 modular control system and have a nominal travel range of 31 mm per axis. The Wolter mirror focuses the incident soft X-rays, and the piezoelectric stages raster the sample surface through the focused beam. A 3D CAD model provides an overview of the main chamber [Fig. 3(*f*)]. The relative arrangement of the sample holder, piezoelectric scanning stage, vacuum-compatible robot, and Wolter mirror is shown in the photograph in Fig. 3(*g*).

In the demonstrations reported here, the drain current from the samples mounted on the Al plate was collected through the sample-holder assembly and the electrically isolated receiver. A dedicated signal line connected to the receiver carried the current to a Keithley Model 6514 programmable electrometer, and this measured current was used as the XAS signal. The receiver was electrically isolated from the piezoelectric scanning stages using ceramic screws, thereby providing an independent current-collection path. An upstream Au mesh was used to monitor the incident photon intensity, and the final XAS signal was normalized based on the Au-mesh drain current.

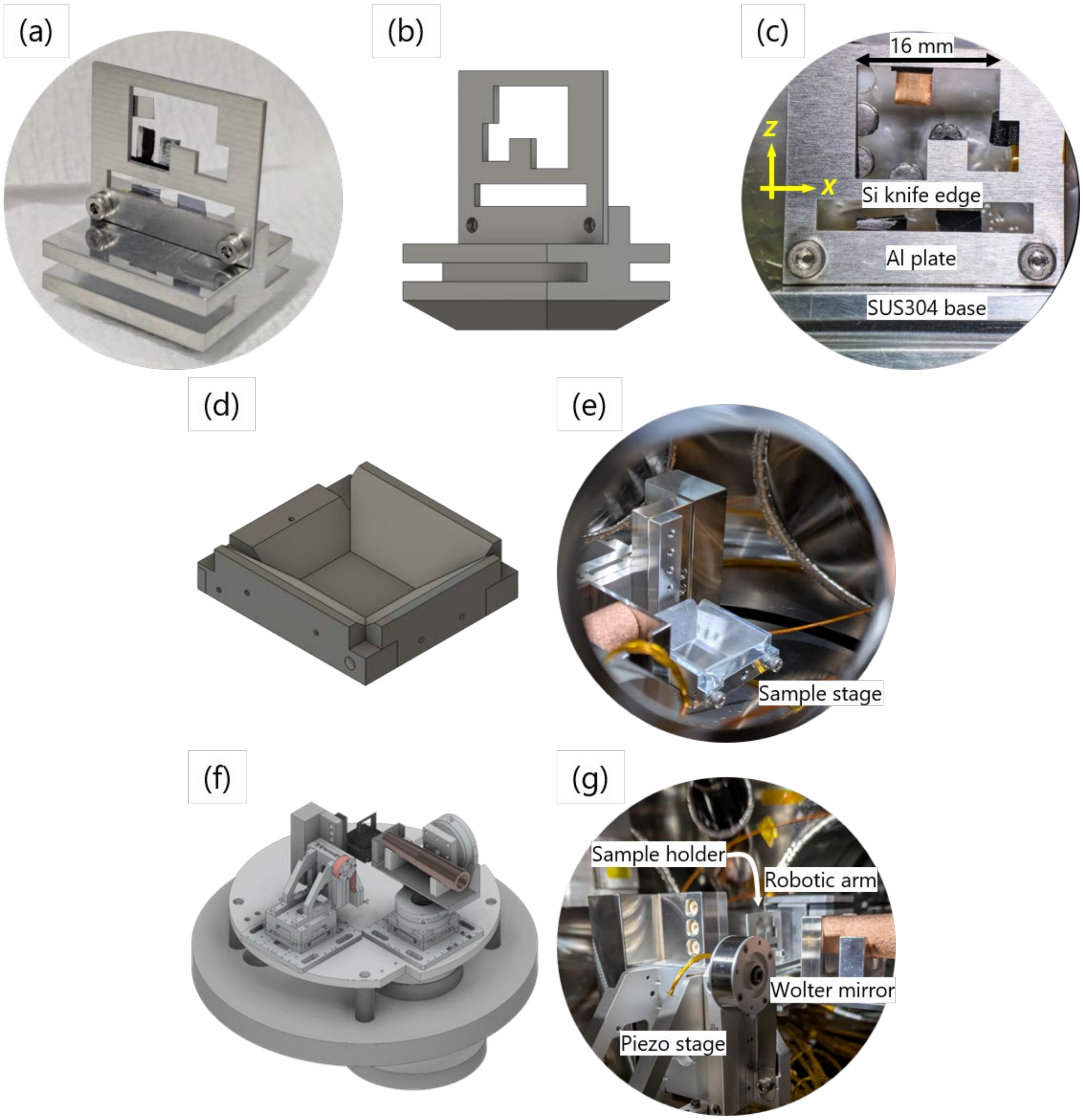

**Figure 3** Sample-holder assembly and scanning station: (*a*) photograph of the assembled sample holder; (*b*) 3D CAD model showing the Al sample plate and SUS304 holder base; (*c*) close-up photograph of the holder, showing the 16 mm opening, scan directions, Si knife edges, and mounting of several samples; (*d*) 3D CAD model of the tapered receiver; (*e*) photograph of the sample stage installed in the main chamber; (*f*) overview 3D CAD model of the main chamber; and (*g*) photograph of the sample holder, piezoelectric scanning stage, vacuum-compatible robot, and Wolter mirror.

### 2.2. Control interface and sequence management

The control panel (TOYAMA Co. Ltd) displays the real-time status of the acrylic box, transfer chamber, UFO chamber, main chamber, and associated gate valves. It also provides manual controls for opening and closing the gate valves and for operating the venting and evacuation functions [Fig. 1(*b*)]. During automated operation, commands are transmitted from the control PC to the control panel through a communication interface [Fig. 1(*c*)]. The sequence executed on the PC relies on flag-based state management, where flags represent the confirmed states of the chambers, valves, robots, and measurement stage, and each operation is initiated only when the required flags have been satisfied.

### 2.3. Beamline optical layout and Wolter-mirror geometry

Fig. 4(*a*) schematically shows the optical path from the APPLE-II undulator at NanoTerasu BL08U to the sample surface. The beam passes through the front-end slit, M0 and M1 mirrors, the plane grating (Gr), the M2 plane mirror, the M3 mirror, and the exit slit (S2). M2 collimates the beam diffracted by Gr. Beyond S2, the monochromatized beam travels along a straight path past an Au mesh, which monitors incident photon intensity, to the Wolter mirror manufactured by Natsume Optical Corporation. Fig. 4(*a*) indicates approximately 16 m from S2 to the Wolter mirror and 40 mm from the mirror to the sample surface. The Wolter mirror comprises ellipsoidal and hyperboloidal reflecting surfaces and focuses the beam through two successive grazing-incidence reflections. Its nominal design focal sizes are 500 nm horizontally and 100 nm vertically under the specified optical conditions.

The arrangement of the Wolter mirror and sample-scanning stage is shown in Fig. 4(*b*). The mirror stage provides three alignment degrees of freedom: translation along the incident-beam direction ($Y$), pitch, and yaw. Longitudinal translation adjusts the focal position along the $Y$ direction, whereas pitch and yaw align the focused beam with the sample position. The modular SmarAct stage architecture allows horizontal and vertical translation units to be incorporated when further alignment flexibility or improved focusing stability is required. The sample-holder receiver is mechanically registered to the nominal focal position and mounted on piezoelectric stages that provide two translational degrees of freedom along the horizontal ($X$) and vertical ($Z$) directions. During spatially resolved measurements, the sample surface is rastered through the fixed focal spot.

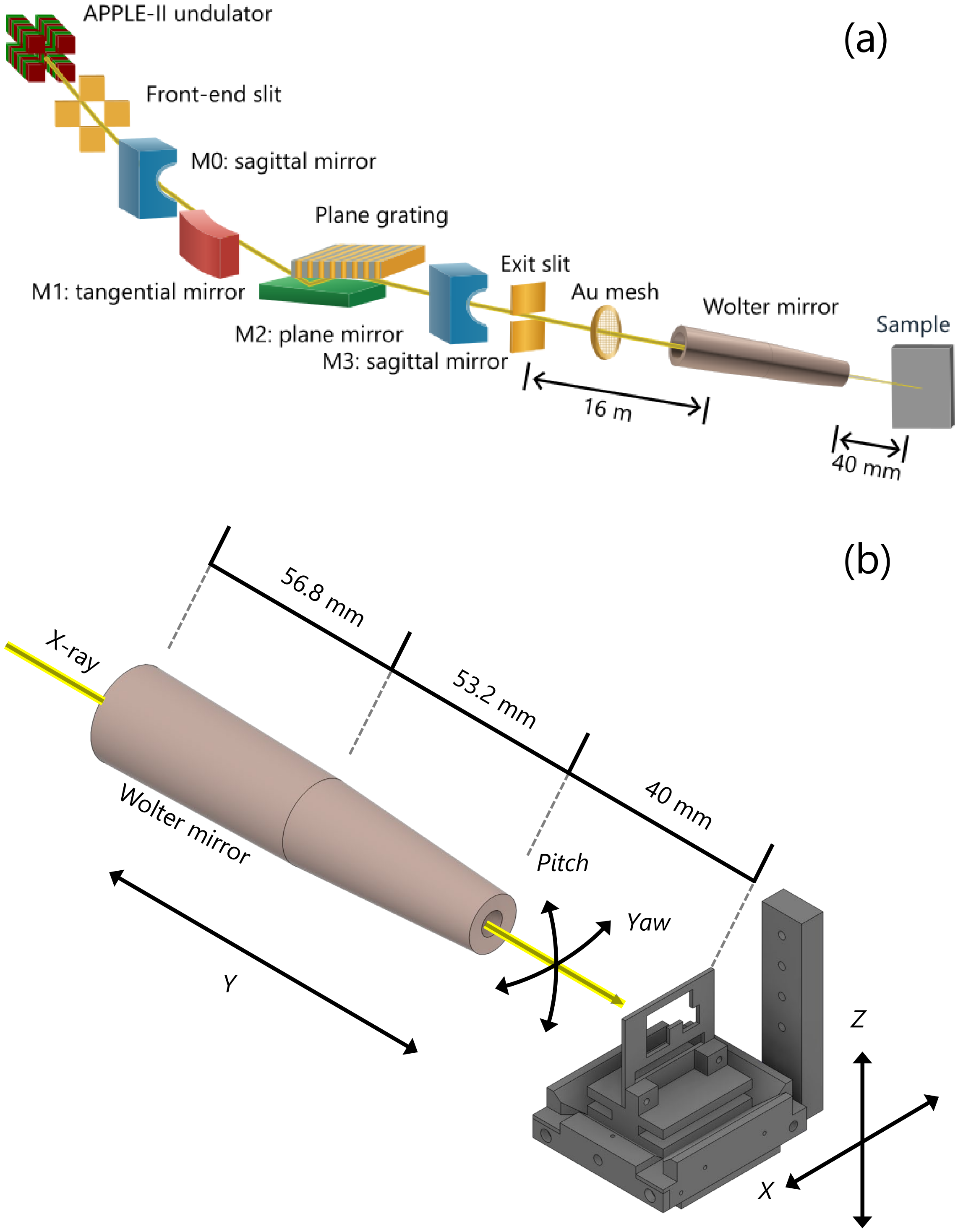

**Figure 4** Schematic optical layout and scanning geometry at the PIONEER end-station. (*a*) Optical path from the APPLE-II undulator to the sample surface, including the beamline mirrors (M0-M3), plane grating, an exit slit, a Au mesh and Wolter mirror. The indicated distances are approximately 16 m from the exit slit to the Wolter mirror and 40 mm from the mirror to the sample surface. (*b*) Enlarged schematic of the Wolter mirror and sample surface. Its stage provides translation along the X-ray direction, pitch and yaw; the sample stage provides *X* and *Z* translation.

### 3. Performance

### 3.1. Beam size and transfer repeatability

The focused-beam dimensions were evaluated using a Si knife edge. At a photon energy of 800 eV, the edge was scanned through the fixed focal spot independently along the *X* and *Z* directions, using a step size of 1 µm and a dwell time of 500 ms per point. Each knife-edge profile was differentiated, and the resulting derivative was fitted with a Gaussian function. The full widths at half-maximum (FWHMs) of the fitted derivatives were 7.5 µm in *X* and 5.3 µm in *Z* (Fig. 5).

The illuminated region of the Wolter mirror, possible mounting-related or gravity-induced deformation, and mechanical vibration of the chamber may contribute to the measured focal-spot size. When a smaller focal spot is required, the focusing conditions can be further optimized upon request. The upstream four-quadrant slit can be adjusted to shape the incident beam and control its footprint on the mirror. The modular stage architecture also allows horizontal and vertical translation units to be added to the existing longitudinal-translation, pitch, and yaw stages without redesigning the end-station. These additional axes would provide greater flexibility to position the mirror relative to the incident beam, potentially allowing illumination of a region with smaller local figure errors while reducing the range of adjustments required for the upstream M3 mirror.

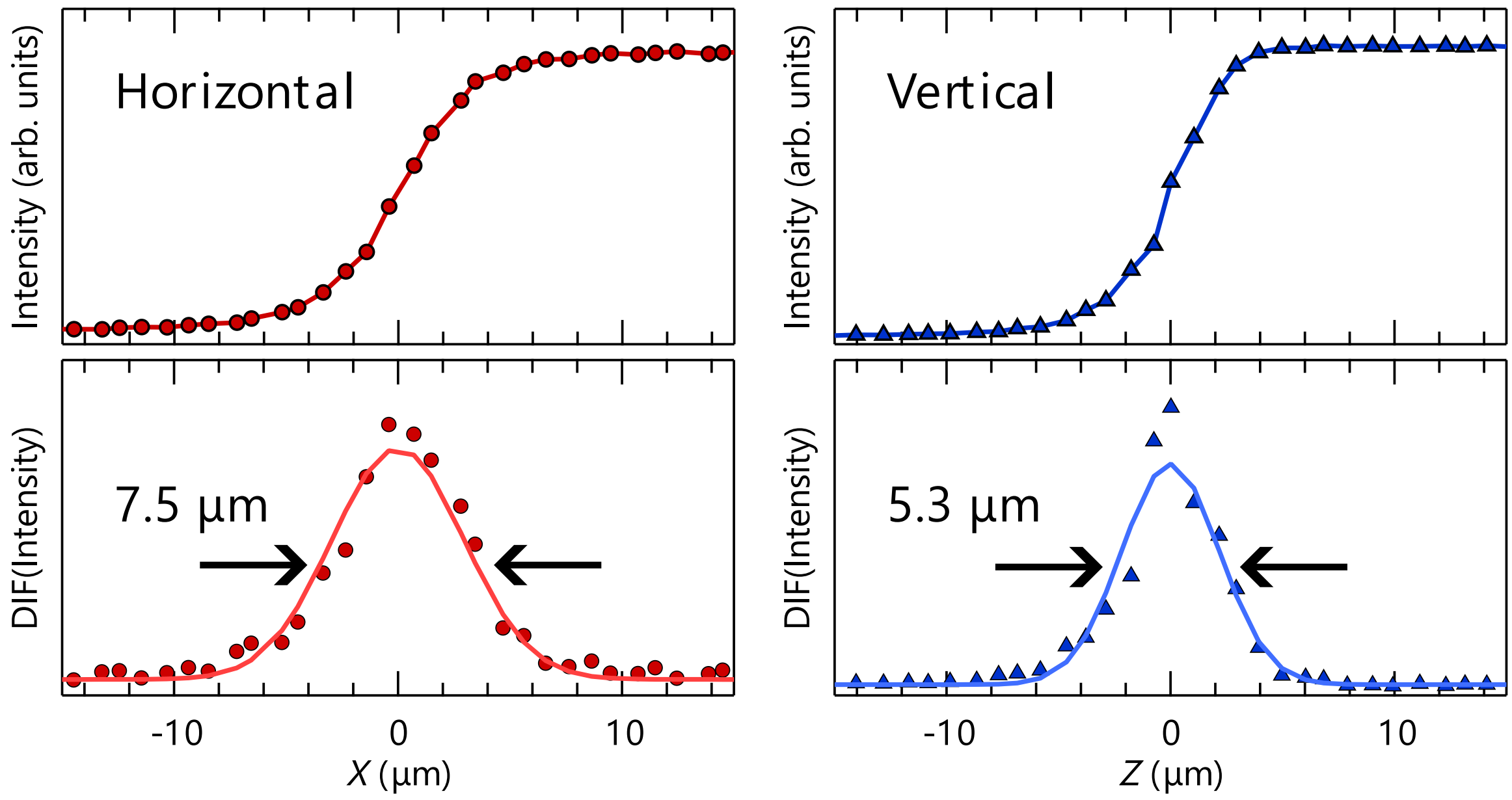

**Figure 5** Determination of the focused-beam sizes using a Si knife edge: upper panels show knife-edge profiles measured along the *X* (left) and *Z* (right) directions; lower panels show the corresponding differentiated profiles, where the circle and triangle markers represent the experimental data, and the solid curves are Gaussian fits (the resulting FWHMs are 7.5 µm in *X* and 5.3 µm in *Z*).

To evaluate the positional reproducibility of the robotic transfer, the same sample-holder assembly was subjected to three successive, complete removal-and-return cycles and remeasured after each cycle. Fig. 6 shows that the fitted edge position varied by approximately ±0.3 μm in *X* and approximately ±0.1 μm in *Z*, while the FWHM remained close to its initial value.

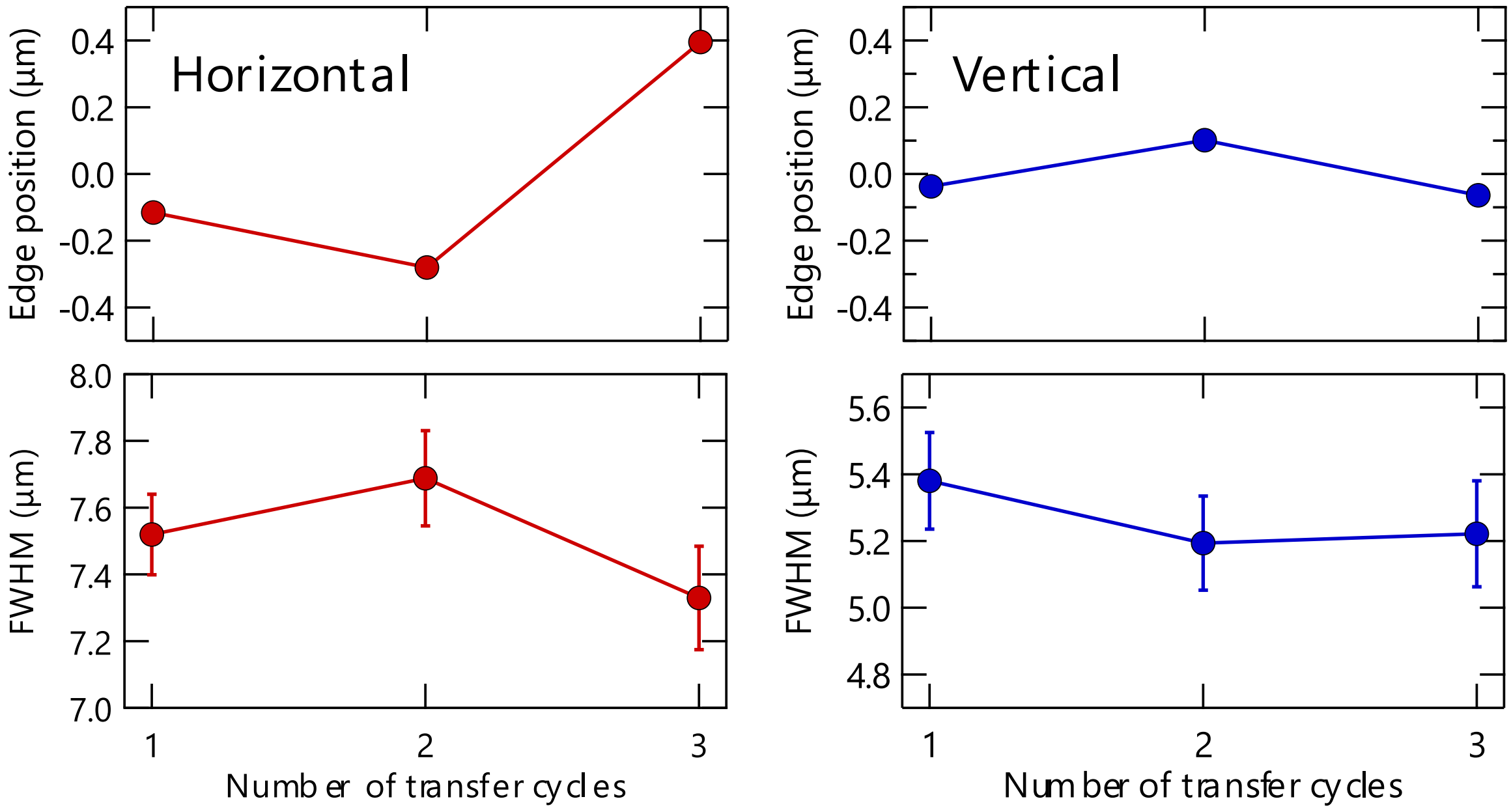

**Figure 6** Positional reproducibility evaluated over three complete sample-transfer cycles: left-hand plots show the results in the horizontal direction, and right-hand plots show those in the vertical direction; upper plots show the fitted edge-position shifts relative to their respective mean values, and lower plots show the fitted FWHMs (error bars represent one standard deviation).

The horizontal direction was selected for the reference test because positional shifts associated with possible sliding of the sample holder at the receiver interface were expected to occur predominantly along this axis. Four consecutive *X*-direction knife-edge scans were acquired without robotic transfer to evaluate the intrinsic repeatability of the scanning stage and acquisition sequence. Fig. 7 shows that the fitted edge position varied within approximately ±0.2 μm, whereas the FWHM remained nearly constant over the four consecutive scans. The horizontal variations observed with and without sample transfer were both substantially smaller than the measured horizontal beam size. Their similar magnitudes indicate that the additional horizontal variation introduced by robotic handling was of the same order as the intrinsic scan-to-scan variation under the present experimental conditions.

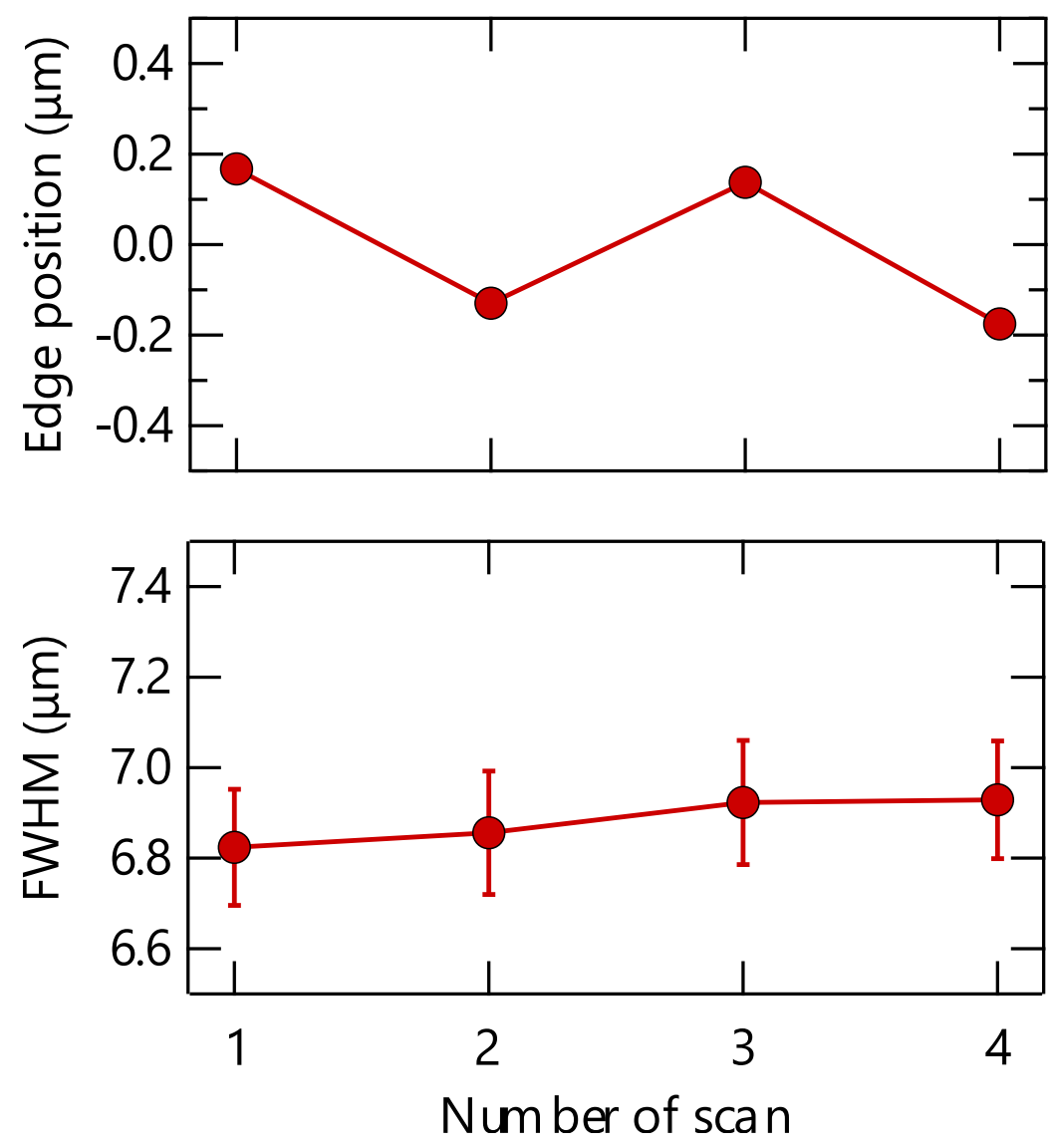

**Figure 7** Repeatability of four consecutive horizontal Si knife-edge scans acquired without robotic sample transfer: upper plot shows the fitted edge-position shift relative to the mean position, and the lower plot shows the fitted FWHM (error bars represent one standard deviation).

### 3.2. Scanning XAS demonstration with a Ni mesh

A Ni mesh was used as a patterned test sample to validate the system. The wire width was approximately 150 µm, and the mesh was mounted on the Al sample plate using conductive carbon tape. Fig. 8 shows the representative Ni *L*-edge spectrum, the measurement geometry, and the spatially resolved measurement of the mesh. The spectrum acquired after automated sample transfer exhibited a high signal-to-noise ratio, with the characteristic Ni $L_3$ and $L_2$ features clearly resolved at approximately 853 and 870 eV, respectively. The photon energy selected for imaging is indicated by the arrow in Fig. 8. The sample-current map acquired at 853 eV covered an area of 300 µm × 300 µm using step sizes of 4 µm in both the *X* and *Z* directions and a dwell time of 100 ms per point. The horizontal and vertical profiles extracted along the green lines show the intensity variations between the Ni wires and mesh openings. The total acquisition time was approximately 30 min. The resulting map reproduced the characteristic geometry of the Ni mesh, demonstrating that the PIONEER system enables the acquisition of high-quality XAS spectra and spatially resolved images following automated air-to-UHV sample transfer.

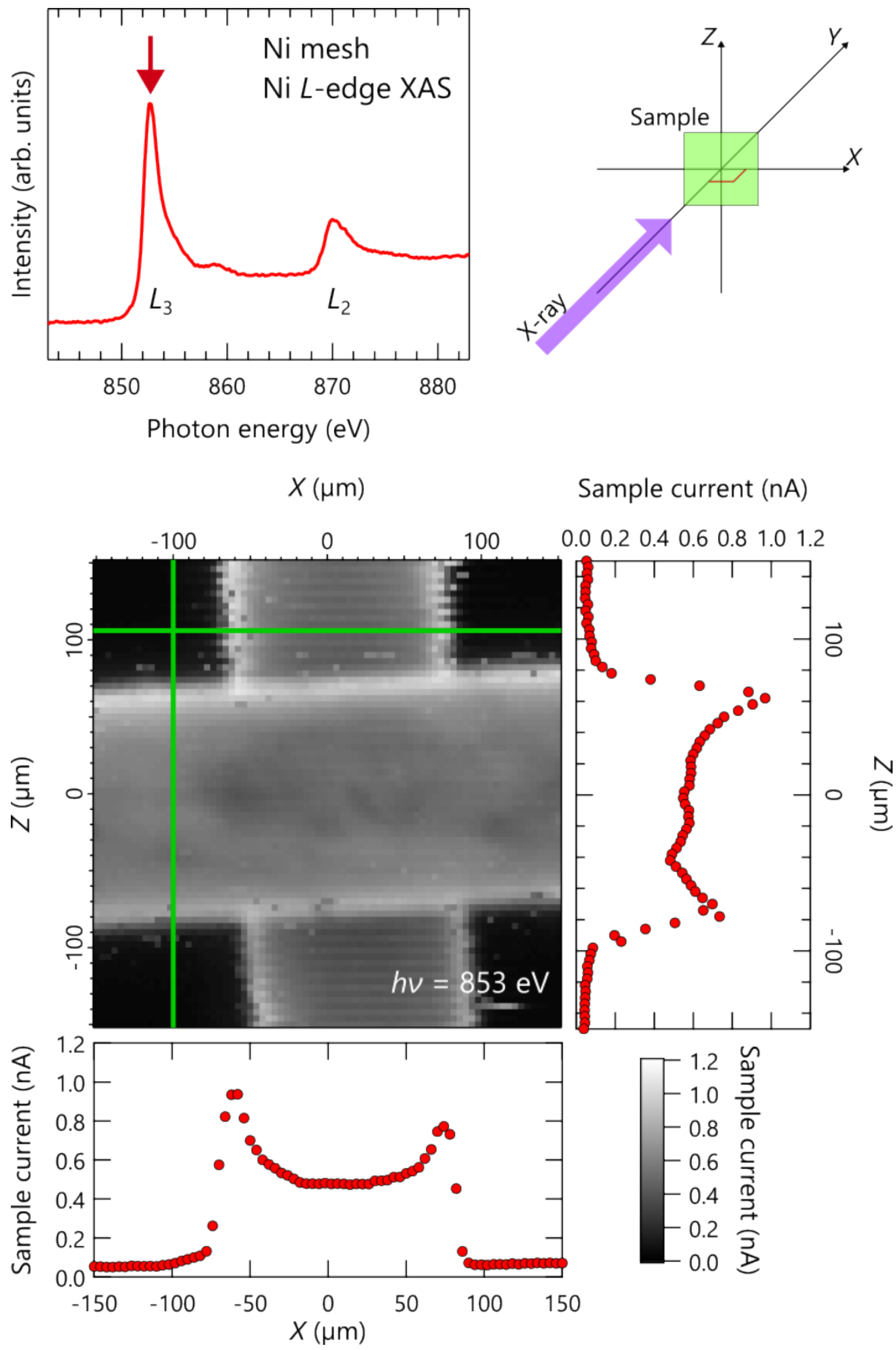

**Figure 8** Spatially resolved Ni *L*-edge measurement of a Ni mesh: the upper-left plot shows the representative Ni *L*-edge spectrum**,** where the arrow indicates the photon energy of 853 eV used for chemical mapping; the upper-right schematic shows the measurement geometry and coordinate system; the chemical map was acquired over an area of 300 μm × 300 μm using step sizes of 4 μm in both the *X* and *Z* directions, where the green horizontal and vertical lines indicate the positions from which the corresponding line profiles were extracted; and the line profiles show the intensity variations across the Ni wires and mesh openings.

Following automated sample transfer and operator registration of the measurement region, XAS measurements were performed on an 11 × 11 spatial grids with a spacing of 10 μm within the region indicated by the yellow square in the inset of Fig. 9. The Ni *L*-edge and O *K*-edge spectra were acquired in separate, sequential photon-energy scans at each of the 121 positions, yielding a total of 242 spatially resolved spectra. An energy step of 0.2 eV was used for both absorption edges. For both energy regions, the sample current was normalized by the drain current from the upstream Au mesh. The Ni *L*-edge spectra exhibited clearly resolved $L_3$ and $L_2$ features, whereas the O *K*-edge spectra showed position-dependent variations in spectral intensity and line shape. The complete sequence, including automated air-to-UHV sample transfer and recipe-based scanning measurements, was completed in approximately 12 h. Following registration of the measurement region and acquisition recipe, the 242 spectra were acquired without further routine intervention by on-site beamline staff. The resulting spectral data and associated metadata were automatically transferred to the data-management system and registered in the linked database. The measurement records can subsequently be retrieved for remote examinations and data analysis.

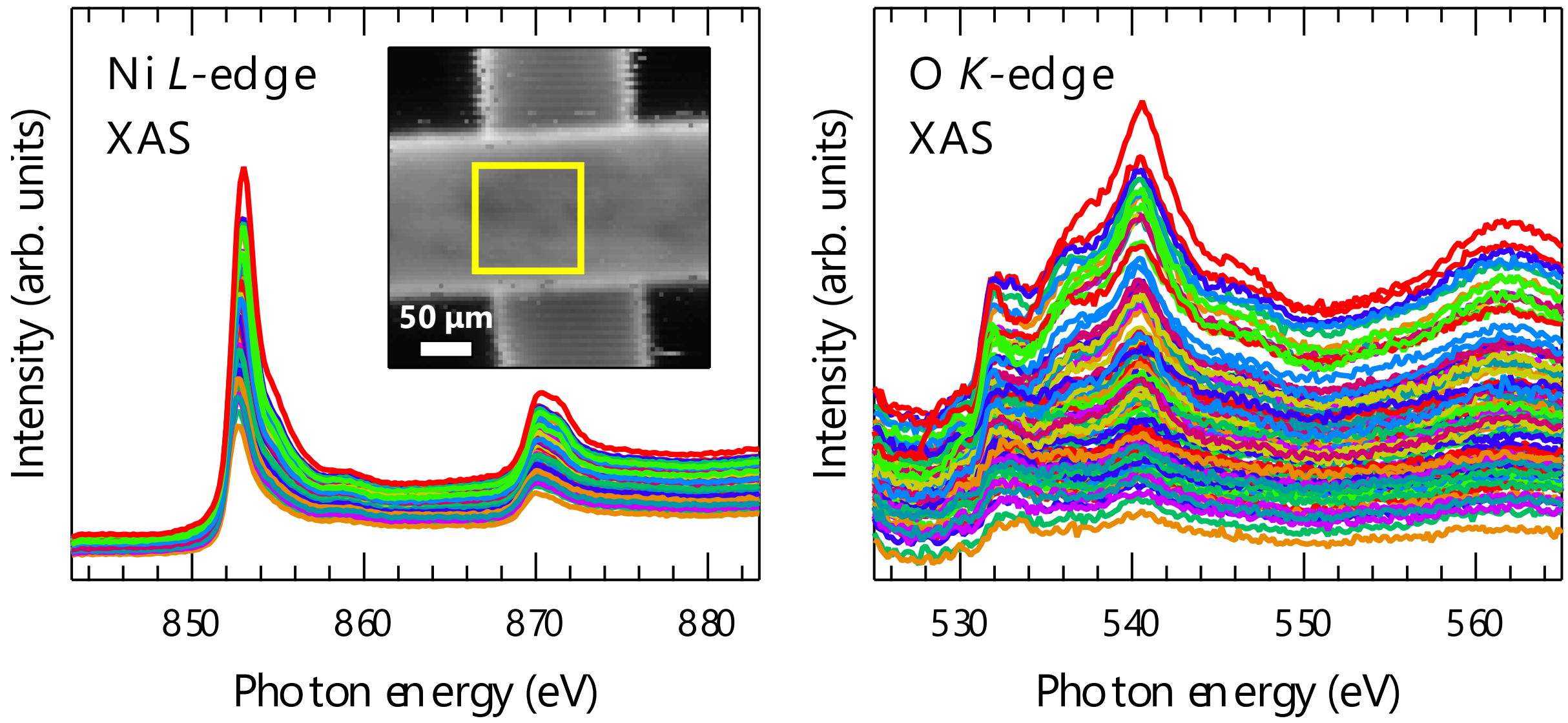

**Figure 9** Spatially resolved Ni *L*-edge and O *K*-edge XAS spectra acquired from the Ni mesh following automated sample transfer: left- and right-hand plots show the Ni *L*-edge and O *K*-edge spectral series, respectively; the inset shows the sample-current image acquired at 853 eV, and the yellow square indicates the region in which the spatially resolved spectra were collected. Each colored curve represents a spectrum acquired at a different position within the selected region. The sample current was normalized by the drain current from the upstream Au mesh.

### 3.3. Machine-learning analysis

The 121 Ni *L*-edge spectra acquired on the 11 × 11 spatial grids were analyzed using four dimensionality-reduction methods: PCA, MDS, t-SNE, and UMAP. These analytical functions are incorporated in our home-made software that is accessible by users. Each spectrum was treated as a multidimensional vector defined by its intensity values over the measured photon-energy range. The analysis settings and hyperparameters are summarized in Table S2 of the supporting information. Fig. 10 compares the low-dimensional representations obtained using the four methods. Each point represents a Ni *L*-edge spectrum acquired at a single spatial position. All four methods captured variations within the spectral dataset, but UMAP provided the clearest separation of spectra with distinct characteristics.

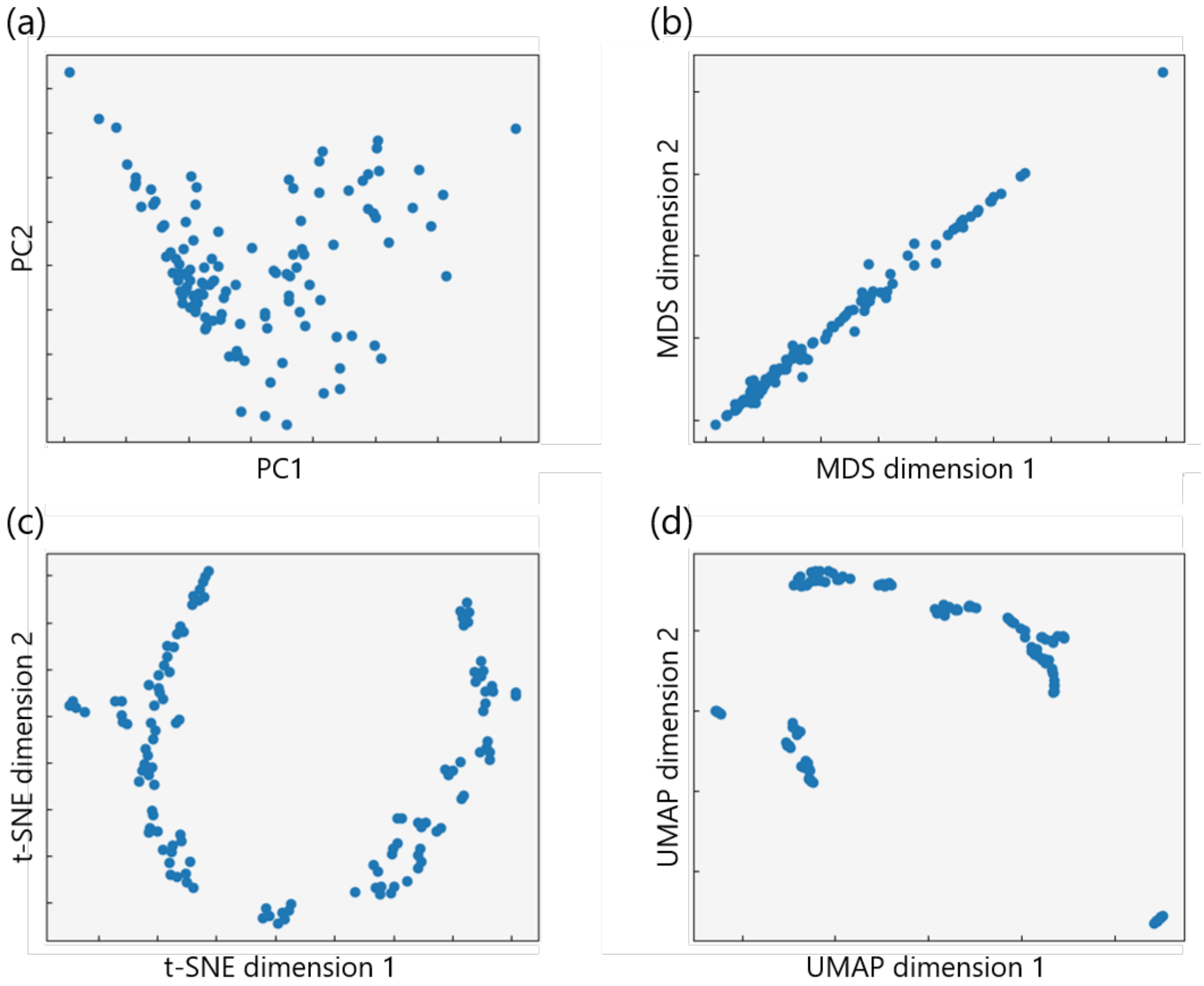

**Figure 10** Comparison of four dimensionality-reduction methods [(*a*) PCA, (*b*) MDS, (*c*) t-SNE, and (*d*) UMAP] applied to the spatially resolved Ni *L*-edge XAS dataset: each point represents one of the 121 spectra acquired on the 11 × 11 spatial grids. Among the four methods, UMAP provided the clearest separation of spectra with different Ni *L*-edge characteristics.

The *k*-means clustering algorithm was subsequently applied to the two-dimensional coordinates obtained by UMAP. The number of clusters was set to four based on an elbow analysis of the within-cluster sum of squares for *k* = 1–10 (Fig. S1 of the supporting information). The 121 spectra were thus classified into four clusters [Fig. 11(*a*)]. The resulting cluster labels were mapped back onto the corresponding positions of the 11 × 11 measurement grid to visualize their spatial distribution [Fig. 11(*b*)]. An average Ni *L*-edge spectrum was calculated from all spectra assigned to each cluster [Fig. 11(*c*)]. The average spectra exhibited characteristic differences in the Ni $L_2$-edge structure, as indicated by the arrows in Fig. 11(*c*). Based on these spectral features, the four clusters were assigned to different Ni oxidation states (Ohldag *et al.*, 2001; Regan *et al.*, 2001). Projecting the oxidation-state assignments back to the original measurement coordinates produced a spatial map of the Ni chemical states within the analyzed region.

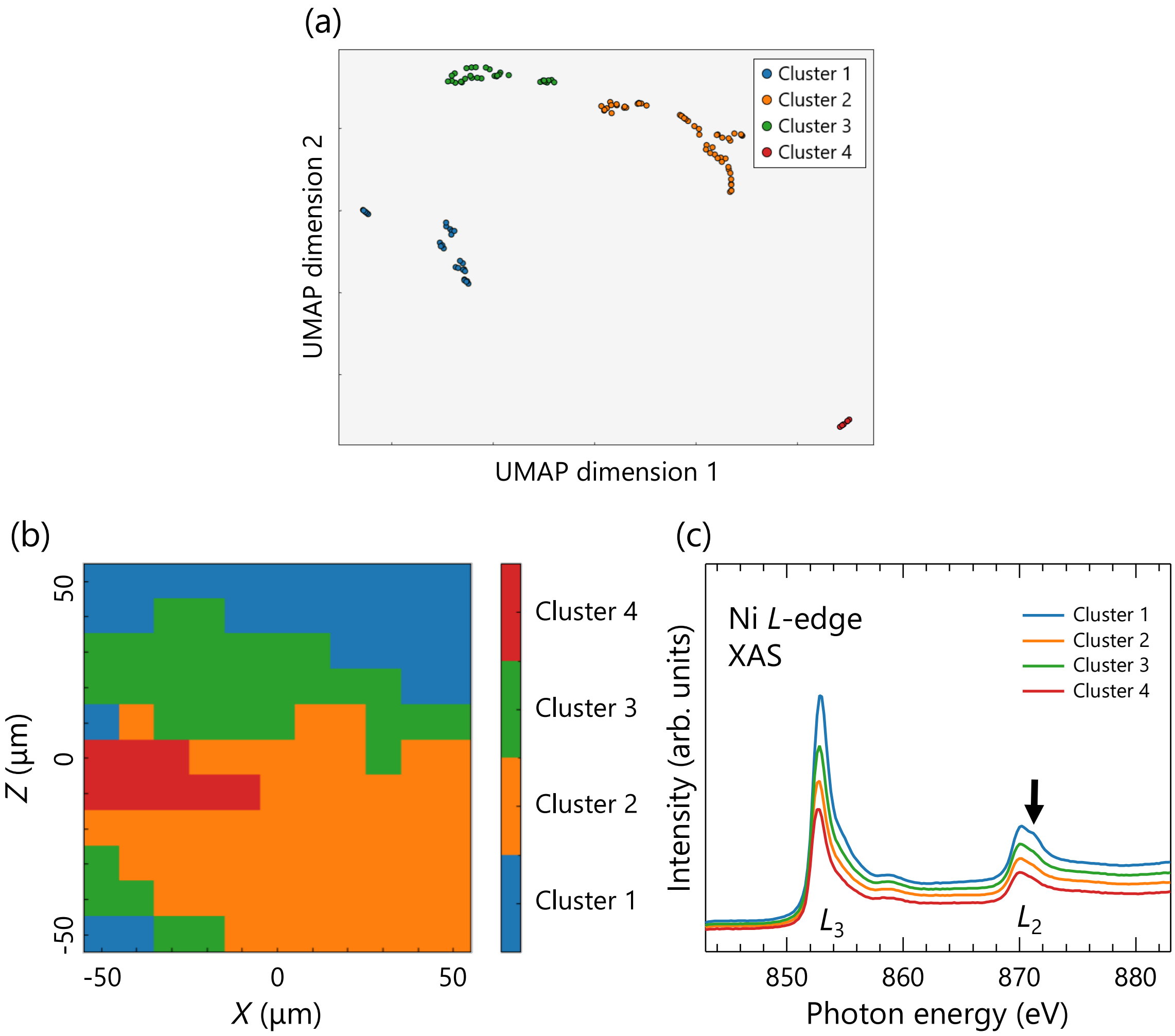

**Figure 11** Classification and spatial visualization of Ni oxidation states using UMAP and *k*-means clustering: (*a*) two-dimensional UMAP representation of the 121 Ni *L*-edge spectra, where *k*-means clustering applied to the UMAP coordinates classified the spectra into four clusters; (*b*) spatial distribution of the four clusters reconstructed on the original 11 × 11 measurement grid; (*c*) average Ni *L*-edge spectra calculated from all spectra assigned to each cluster, where the arrows indicate characteristic differences in the Ni $L_2$-edge structure used to assign the clusters to different Ni oxidation states. The same colors represent the corresponding clusters in all panels.

Because the classification used the complete Ni *L*-edge line shape, the spatial variation in oxidation state could be visualized without applying individual peak fitting or manual feature extraction to every spectrum. The resulting low-dimensional coordinates, cluster assignments, cluster-averaged spectra, and final oxidation-state map were registered in the project database alongside the original XAS dataset and associated measurement and analysis metadata. These results demonstrate that pairing UMAP with *k*-means clustering enables efficient, automated classification and spatial visualization of Ni oxidation states in large-scale scanning XAS datasets.

## 4. Human–robot collaboration

The modular architecture of the PIONEER system at NanoTerasu BL08U provides a foundation for future autonomous soft X-ray experiments. PIONEER stores acquired XAS data in cloud storage and automatically registers the associated measurement conditions and beamline metadata in a linked database, allowing experimental records to accumulate across measurements. The available space in the atmospheric enclosure and the sample-holder interface could accommodate automated synthesis or *operando* sample-preparation modules. In a future iterative workflow, newly acquired measurements, together with accumulated data, could inform the selection of subsequent synthesis conditions according to a researcher-defined objective. Bayesian optimization provides one possible strategy for this process, building on our earlier demonstration using a robotic solvent-mixing system (Tang *et al.*, 2025). Integration of these functions could establish a closed synthesis–transfer–measurement–analysis loop and extend the end-station from systematic characterization toward autonomous, data-driven exploration of the relationships between synthesis conditions and electronic states during beamtime.

The developed system also changes the division of work between automation and people at the beamline, as illustrated in Fig. 12. Once the target region and acquisition recipe have been registered, routine operations—including air-to-UHV sample transfer, vacuum and gate-valve control, scanning measurements, data recording, metadata registration and sample return—are performed automatically. The unstaffed end-station therefore shifts synchrotron user experiments away from labor-intensive manual operation toward automated, objective-driven data acquisition. Researchers and beamline staff remain responsible for defining scientific objectives, designing experiments, interpreting the resulting data and making scientific decisions. By reducing repetitive operational tasks and improving the effective use of allocated beamtime, PIONEER allows researchers, beamline scientists and technicians to devote more time to sophisticated experimental design, scientific interpretation, instrument development and other creative scientific activities (Wada *et al.*, 2025; Miyamoto *et al.*, 2025; Horio *et al.*, 2025; Murano *et al.*, 2026).

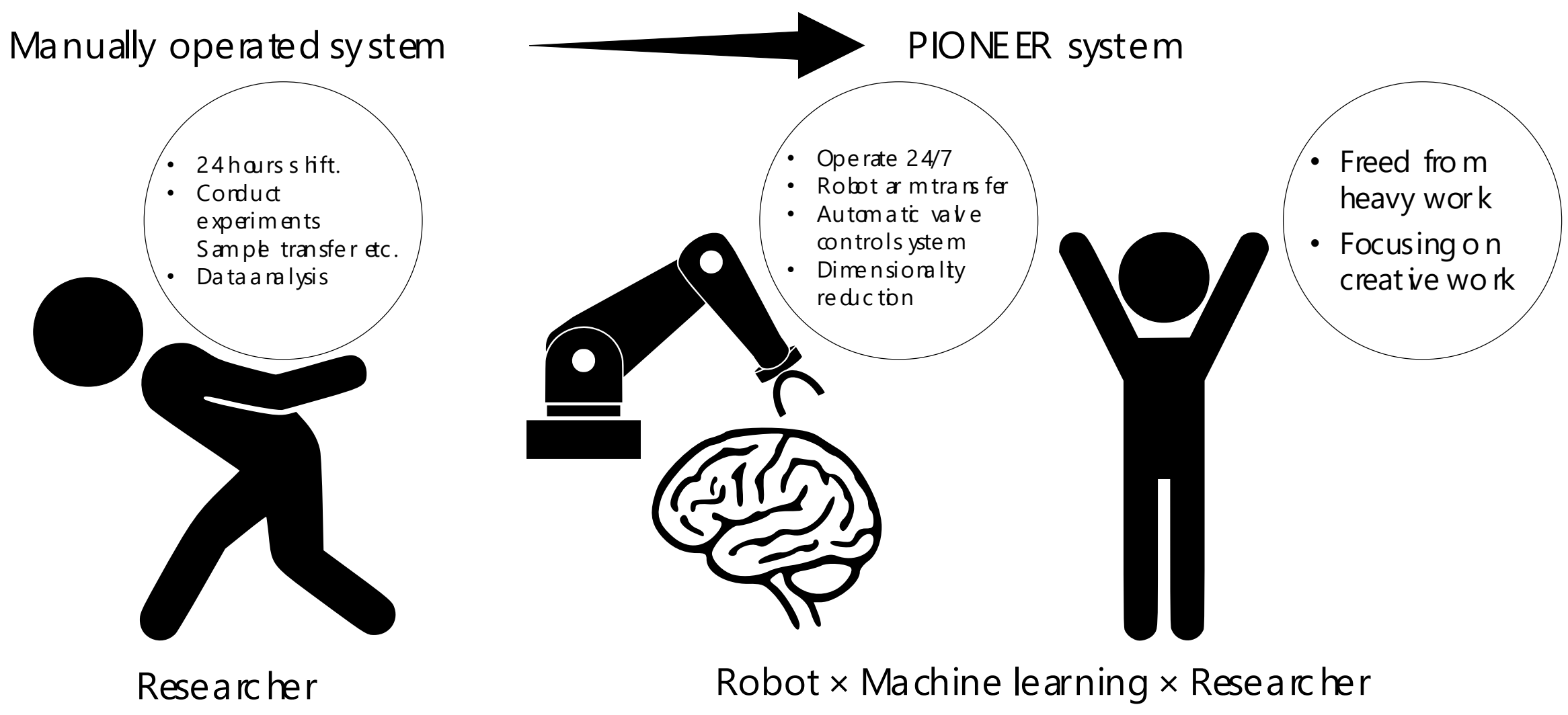

**Figure 12** Conceptual comparison of conventional operator-intensive synchrotron experiments with the automated PIONEER workflow. In conventional scanning soft XAS experiments, researchers and beamline staff must remain involved in repetitive operations such as sample transfer, instrument control, and data acquisition. The PIONEER system combines robotic air-to-UHV sample transfer, automated vacuum and gate-valve control, predefined scanning measurements, and dimensionality-reduction analysis. After registration of the measurement conditions, the system supports extended measurements with reduced routine intervention, allowing researchers and beamline staff to focus on experimental design, data interpretation, and other scientific and creative activities.

## 5. Summary

We developed the PIONEER system, an integrated end-station engineered for automated air-to-UHV sample transfer and scanning soft XAS using a micro-focused beam at the NanoTerasu BL08U. The system coordinates atmospheric and vacuum-compatible robots, automated evacuation and venting of the transfer chamber, gate-valve sequencing, sample positioning, spatial and photon-energy scanning, data recording, metadata registration, and sample return within a single workflow. After the operator identifies the target sample and registers the measurement region and acquisition parameters, the PIONEER system executes the measurement and returns the sample holder without further routine intervention.

The efficacy of this integrated workflow was validated through spatially resolved Ni *L*-edge and O *K*-edge measurements of a Ni mesh. XAS spectra were acquired at 121 positions on an 11 × 11 spatial grid, yielding a total of 242 spectra in approximately 12 h. Meanwhile, all spectral data and associated experimental metadata were autonomously archived. UMAP followed by *k*-means clustering classified the Ni *L*-edge spectra into four groups with distinct spectral features, which were assigned to different Ni oxidation states and mapped onto the measurement grids. Together, these results demonstrate that the PIONEER system can perform extended micro-focused scanning XAS measurements and systematically generate and analyze large-scale datasets under consistent experimental conditions.

**Acknowledgements** This work was supported by JST CREST, Japan (Grant No. JPMJCR21O4). The authors gratefully acknowledge the Photon Science Innovation Center (PhoSIC) and the Japan Synchrotron Radiation Research Institute (JASRI) for their technical support in the development and operation of the PIONEER system. The authors also thank Robert Ireland, PhD, from Edanz (https://jp.edanz.com/ac) for editing a draft of this manuscript.

**Funding information** This work was supported by JST CREST, Japan (Grant No. JPMJCR21O4).

**Conflicts of interest** The authors declare that they have no conflicts of interest.

**Data availability** The data supporting the findings of this study are available from the corresponding author upon reasonable request.

## Appendix A. Automated transfer movie

Movie 1 Automated transfer of a sample holder from the atmospheric sample bank to the UHV measurement stage, followed by its return to the bank after measurement.